\documentclass[paper,nofootinbib] {revtex4}

\usepackage{graphicx}
\usepackage{epsfig}
\usepackage{fancyhdr}
\usepackage{empheq}
\usepackage{mathbbol}
\usepackage{wasysym}
\usepackage{pstricks}
\usepackage{color}
\usepackage{bbm}
\def\mol{D^0\bar D^{*0}}
\def\en{{\cal E}}

\begin{document}

\title{Is the $X(3872)$ Production Cross Section at Tevatron Compatible with a\\ 
 Hadron Molecule Interpretation?}

%

\author{C Bignamini$^{\dag,\ocircle}$, B Grinstein$^{*}$, F Piccinini$^{\dag}$, AD Polosa$^\P$} 
\author{C Sabelli$^{\ddag,\P}$}
\affiliation{
$^\dag$ INFN Pavia, Via A. Bassi 6, Pavia, I-27100, Italy\\
$^\ocircle$ Institut f\"ur Theoretische Physik, Universit\"at Karlsruhe, Karlsruhe, D-76128 Germany\\
$^*$University of California, San Diego, Department of Physics, La Jolla, CA 92093-0315, USA\\
$^\P$INFN Roma, Piazzale A. Moro 2, Roma, I-00185, Italy\\
$^\ddag$Department of Physics, Universit\`a di Roma, `La Sapienza', Piazzale A. Moro 2, Roma, I-00185, Italy}

\begin{abstract}
The $X(3872)$ is universally accepted to be an exotic hadron. In this letter we assume that the $X(3872)$ is a $\mol$ molecule,
as claimed by many authors, and attempt an estimate of its  prompt production cross section at Tevatron. 
A comparison with CDF data allows to draw rather compelling quantitative conclusions about
this statement. 
\\ \\
PACS: 12.39.-x, 12.39.Mk, 13.75.-n
\end{abstract}

\maketitle

\thispagestyle{fancy}

{\bf \emph{Introduction}}.
Since the discovery of the $X(3872)$ resonance by Belle~\cite{bellex} it was soon realized that, despite its prominent decay mode were into $J/\psi\rho$,
this particle could not have  been identified as a standard charmonium excitation~\cite{quigg}.  The mass of the $X$ being so close to the $\mol$ 
threshold suggested to many authors that it could be a neat example of a hadron molecule~\cite{mols}. The idea of molecules of hadrons dates back to the work 
by Fermi and Yang~\cite{fermiyang}  in which the pion is interpreted as a nucleon-nucleon bound state. The case of the $X(3872)$, in this respect, is 
rather peculiar though: the $\mol$ molecule constituting the $X(3872)$ would be characterized by an extremely small binding energy.

After Belle, CDF and D0 confirmed the $X(3872)$ in proton-antiproton collisions~\cite{cdfx,d0x}. Later, BaBar~\cite{babarx}  found the $X(3872)$ in the same 
$B$ decay observed at Belle. It seems odd that a such a loosely bound molecule could be produced promptly ({\it i.e.} not from $B$ decay) in a high 
energy hadron collision environment.

This was one of the motivations to consider the possibility that the $X(3872)$ could be, instead of a molecule, a `point-like' hadron resulting from 
the binding of a diquark and an antidiquark~\cite{4q}, following the interpretation proposed by Jaffe and Wilczek~\cite{jw} of pentaquark baryons (antidiquark-antidiquark-quark). 
The drawback of the diquark picture is the proliferation of states predicted  and the little insight on selection rules that could explain why many 
of these states are not yet seen (for a recent account on this aspect see {\it e.g.}~\cite{dfp}).   

In this letter we address the problem of estimating the prompt production cross section of the $X(3872)$ at the Tevatron and compare our results  with the most recent observations by the CDF experiment allowing us to obtain a lower bound for it. We will assume that the $X(3872)$ is a $\mol$ hadron molecule.  For an earlier attempt in this direction~\cite{suzuki}.

We employ standard Monte Carlo tools as Herwig~\cite{herwig} and Pythia~\cite{pythia} to 
compute  the production of $D$ and $D^*$ hadrons in  proton-antiproton collisions at Tevatron energies.
Open charm meson pairs produced are ordered as a function of their relative 3-momentum and of their center of mass momentum.
We select those which  pass the kinematical cuts used in the analysis made by the CDF collaboration.

As we will explain below, we can estimate an {\it upper bound} for the theoretical cross section and a {\it lower bound} for the experimental one.
The comparison of the two should give a qualitative answer  whether the production of $X(3872)$ is exclusively due to the formation of a 
molecular bound-state.
In the following considerations we will not make use of any particular model. The only model dependency  in our calculations is that
hardwired in the hadronization schemes of Pythia and Herwig. Indeed, to add weight to our conclusions, a comparison between the results 
obtained with both MC's will be carried out.

{\bf \emph{CDF data on prompt $X(3872)$ production}}. 
The CDF collaboration soon after the observation of the $X(3872)$ in the $J/\psi \pi^+\pi^-$ channel
measured the fraction of promptly produced $X(3872)$ to be $83.9 \pm 5.3 \% $~\cite{cdf}.
In the same conference note we find the yields of 
$\psi(2S)\to J/\psi \pi^+\pi^-$ and $X(3872)\to J/\psi \pi^+\pi^-$ candidates,
using nearly identical selection criteria (indeed the 
selection differs by just an additional cut on the dipion invariant mass 
for the $X(3872)$ case which is reported to produce little effect on the $X(3872)$ signal),
as well as the measurement of the prompt fraction of $\psi(2S)$ candidates. From these, and 
assuming the same detection efficiency for $\psi(2S)$ and $X(3872)$,
which is presumably true within a factor of two but has never been reported 	
as such by the CDF collaboration, we can roughly estimate, taking from~\cite{pdg} the 
{\cal B}($\psi(2S)\to J/\psi \pi^+\pi^-)$:
\begin{equation}
\frac{\sigma(p\bar p\to X(3872)+{\rm All})_{\rm prompt}\times {\cal B}(X(3872)\to J/\psi \pi^+\pi^-))}
     {\sigma(p\bar p\to \psi(2S)+{\rm All})} = 4.7 \pm 0.8 \%
\end{equation}
where the uncertainty includes statistical and systematic uncertainties but do not 
attempt to gauge the assumption of equal detection efficiency. The acceptance of the $\psi(2S)$ and
$X(3872)$ candidates is not specified in~\cite{cdf}, but from the CDF\ II detector geometry and 
the indicated candidates selection we can conservatively assume that the above ratio applies for 
$p_\perp>5\ {\rm GeV}$ and $|y|<1$. 

To derive an absolute $X(3872)$ cross section, we can use the recently published~\cite{psi2s} 
$\psi(2S)$ Run II prompt differential cross-section measurement integrating from $p_T>5$ and taking
{\cal B}($\psi(2S)\to \mu^+\mu^-$) from~\cite{pdg}:
$\sigma(p\bar p\to \psi(2S)+{\rm All})= 67\pm 9~{\rm nb}$ for
$p_\perp(\psi(2S))>5~{\rm GeV}, |y(\psi(2S))|<0.6$.

 Assuming that both $X(3872)$ and $\psi(2S)$ have the same rapidity distribution in 
the range $|y|<1$, we can finally estimate a lower bound on the prompt prodution cross section of 
$X(3872)$ as:
\begin{equation}
\sigma(p\bar p\to X(3872)+{\rm All})^{\rm min} > \sigma(p\bar p\to X(3872)+{\rm All}) 
  \times {\cal B}(X(3872)\to J/\psi \pi^+\pi^-) = 3.1 \pm 0.7~{\rm nb}  
\end{equation}
in $p_\perp(X)>5~{\rm GeV}, |y(X)|<0.6$. We aim at comparing $\sigma^{\rm min}_{\rm exp}$ with some upper bound on the theoretical determination of the cross section: $\sigma^{\rm max}_{\rm th}$.

{\bf \emph{Estimating an upper bound for $\sigma_{\rm th}$}}. 
Let us suppose that $X(3872)$ is an $S$-wave bound state of two $D$ mesons, namely a $1/\sqrt{2}(D^0\bar D^{*0}+\bar D^0 D^{*0})$ molecule (we will  use the shorthand notation $D^0\bar D^{*0}$). Such a molecule has the correct $1^{++}$ quantum numbers of the $X(3872)$. 
The $X(3872)$ prompt production cross section at the Tevatron could be written as:
\begin{eqnarray}
&&\sigma(p\bar p\to X(3872)) \sim \left| \int d^3 {\bf k} \langle X|D\bar D^{*}({\bf k})\rangle\langle D\bar D^{*}({\bf k})|p\bar p\rangle\right|^2
\simeq \left| \int_{\cal R} d^3 {\bf k} \langle X|D\bar D^{*}({\bf k})\rangle\langle D\bar D^{*}({\bf k})|p\bar p\rangle\right|^2 \nonumber\\
&&\leq \int_{\cal R} d^3 {\bf k} |\psi({\bf k})|^2 \int_{\cal R} d^3 {\bf k} | \langle D\bar D^{*}({\bf k})|p\bar p\rangle |^2
\leq\int_{\cal R} d^3 {\bf k} | \langle D\bar D^{*}({\bf k})|p\bar p\rangle |^2\sim \sigma(p\bar p\to X(3872))^{\rm max}
\label{eq:erre}
\end{eqnarray}
where ${\bf k}$ is the relative 3-momentum between the $D({\bf p_1}),D^*({\bf p_2})$ mesons.
$\psi ({\bf k})=\langle X|D\bar D^{*}({\bf k})\rangle$ is some normalized bound state wave function characterizing the $X(3872)$. ${\cal R}$ is the integration region where $\psi({\bf k})$ is significantly different from zero. The first inequality is the Schwartz inequality: the equal sign holds only when $\psi({\bf k})/\langle D\bar D^{*}({\bf k})|p\bar p\rangle$ is equal to a constant almost everywhere in the integration region~\cite{tit}. The  matrix element 
$\langle D\bar D^{*}({\bf k})|p\bar p\rangle$ can be computed using standard 
matrix-element/hadronization Monte Carlo programs like Herwig and Pythia.
To do so, we require our MC tools to generate $2\to 2$ QCD  events with some loose partonic cuts.  Configurations with one gluon recoiling from a $c\bar c$ pair, are those configuration expected to produce two collinear charm quarks and in turn collinear open charm mesons. The parton shower algorithms in Herwig and Pythia treat properly these configurations at 
low $p_\perp$ whereas they are expected to be less important at  higher $p_\perp$. We will discuss also these processes at the parton level.

In the hadron samples produced by the shower/hadronization algorithm we list the events containing $D^0\bar D^{*0}$
as a function of their center of mass relative momentum. At this level, the only cuts are those on partons: $p_\perp^{\rm part}>2$~GeV and $|y^{\rm part}|<6$.
If more than one $D^0\bar D^{*0}$ pair is found in the event, we select the pair having the smaller relative 3-momentum $k$. 

\begin{figure}
\begin{minipage}[t]{7truecm} 
\centering
\includegraphics[width=7truecm]{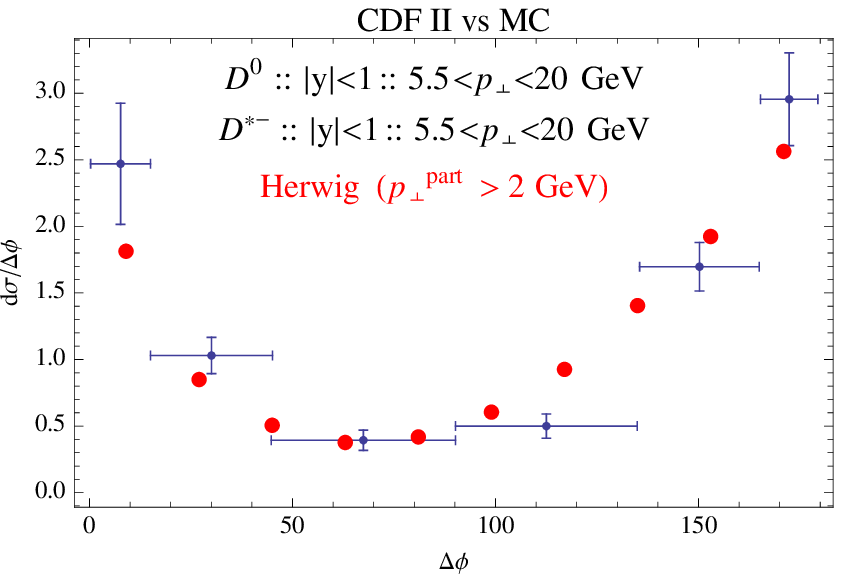}
\caption{The $D^0D^{*-}$ pair cross section as function of $\Delta\phi$ at CDF Run II. The transverse momentum, $p_\perp$, and rapidity, $y$,  ranges 
are indicated. Data points with error bars,  are compared to the leading order  event generator  Herwig. 
The cuts on parton generation are $p_\perp^{\rm part}>2$~GeV and $|y^{{\rm part}}|<6$. We 
have checked that the dependency on these cuts is not significative. 
We find that we have to rescale the Herwig cross section values by a  
factor  $K_{\rm Herwig}\simeq 1.8$ to best fit the data on 
open charm production. }
\label{fig:tune}
\end{minipage}
\hspace{1truecm} 
\begin{minipage}[t]{7truecm}
\centering
\includegraphics[width=7truecm]{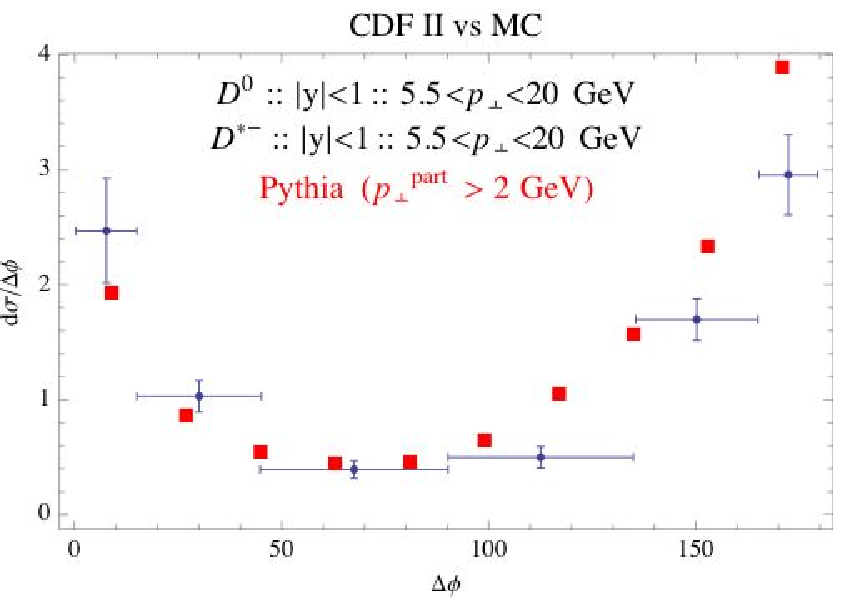}
\caption{The same as in Fig.~\ref{fig:tune} but using Pythia. 
We find that we have to rescale the Pythia cross sections by a factor  $K_{\rm Pythia}\simeq 0.74 $ to best fit the data on 
open charm production. In both cases the agreement of the Monte Carlo distribution with data is remarkable.}
\label{fig:tune2}
\end{minipage}
\end{figure}

We tune our MC tools on CDF data  on $D^0D^{*-}$ pair production cross section distributions in the $\Delta\phi$ variable, $\phi$ being the
azimuthal angle in the transverse plane to the beam axis $z$; see Fig.~\ref{fig:tune} and~\ref{fig:tune2}.

As for the determination of the region ${\cal R}$ in (\ref{eq:erre}) we estimate it having in mind a naive gaussian ansatz for the bound state wave function.
It is straightforward to estimate the momentum spread of the gaussian by  assuming a strong interaction 
Yukawa potential between the two $D$ mesons. Given that
the binding energy ${\cal E}_0$ is ${\cal E}_0\sim M_X - M_D- M_{D^*}= -0.25\pm 0.40$~MeV
 we find that $r_0\sim 8$~fm ($8.6\pm1.1$~fm) and
applying the (minimal) uncertainty principle relation, we get the gaussian momentum
spread  $\Delta p \sim 12$~MeV. One can  check that changing the value of $g^2/4\pi\sim {\cal O}(10)$ has a small effect on the size of ${\cal R}$.

Given the very small binding energy  we can estimate $k$ to be as large as $k\simeq \sqrt{2\mu (-0.25+0.40)}\simeq 17$~MeV or of the order of the center of mass momentum
$k=\sqrt{\lambda(m_X^2,m_D^2,m_D^{*2})}/2m_X\simeq 27$~MeV. 
Given these considerations,
we can restrict the integration region to  a ball ${\cal R}$ of radius\footnote{Which corresponds to a $k_0$ of the Gaussian at $\sim 27$~MeV and a spread of $+12$~MeV.} $\simeq[0,35]$~MeV.

Since we assume that $D-D^*$ interactions have a range of $\sim 1/m_\pi$, we expect a relative orbital angular momentum $\ell \lesssim k/m_\pi$, i.e., we can only allow $S-$wave resonance scattering. Moreover we expect that $\mol$ is a rather narrow object, its width being almost equal to the width of its $D^{*}$ component: $\Gamma\sim 65$~KeV. This is compatible with the determination of Belle and BaBar which find that the width of the $X(3872)$ in the $J/\psi \rho$ channel is $<2.3$~MeV at $90\%$~C.L..
On the other hand attractive potentials  {\it do not} generate such {\it sharp} resonances in $S$-wave. In higher partial waves 
the centrifugal angular momentum  barrier 
allows the formation of bound metastable states.  Although as the $\mol$ molecule has to be a $1^{++}$ state, we would need the first even parity wave, namely $D$-wave! Indeed in higher partial waves one can estimate the width of the resonance to be  $\Gamma\sim \Delta\en\sim \en_0(ka)^{2\ell-1}$, $a$ being the range of the interaction.  Then, in the $D-$wave case, there could have been a sharp resonance that we {\it do not} expect in the $S-$wave case where we necessarily are.  
Other molecule formation mechanisms under study, namely Feshbach resonances, could explain  the narrowness of theses states~\cite{braatenj}.

{\bf \emph{Results}}. 
As shown in Figs.~\ref{fig:tune} and~\ref{fig:tune2}, we can reproduce rather well the cross section distributions in azimuth intervals $\Delta\phi$ for open charm production at CDF (see for example~\cite{intnotecdf} and the relative CDF internal notes), provided that we adopt some rescaling factors as to get the right normalizations.  

\begin{figure}
\begin{minipage}[t]{7truecm} 
\centering
\includegraphics[width=7cm]{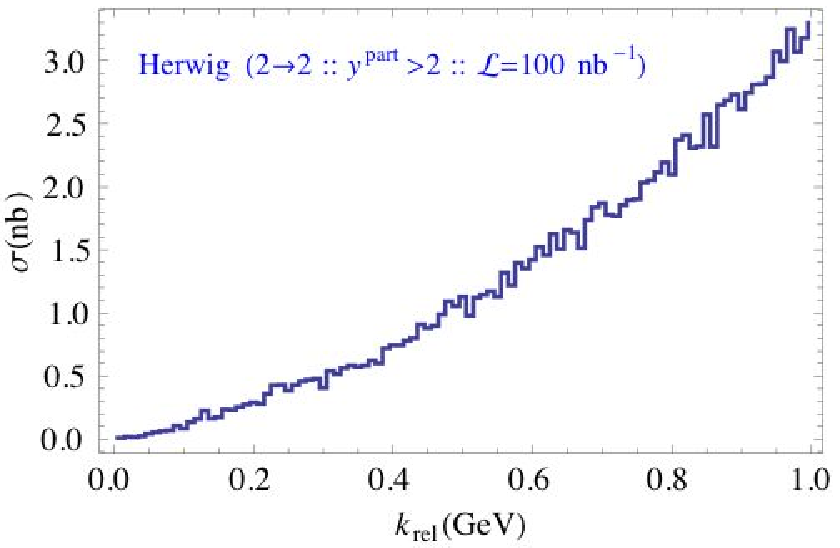}
\caption{The integrated cross section obtained with  Herwig as a function of the center of mass relative momentum of the mesons in the $\mol$ molecule. This plot is obtained after the generation of $55\times 10^{9}$ events with parton cuts $p_{\perp}^{{\rm part}}>2$~GeV and $|y^{{\rm part}}|<6$. The cuts on the final $D$ mesons are such that the molecule produced has a $p_\perp>5$~GeV and $|y|<0.6$.}
\label{herfullc}
\end{minipage}
\hspace{1truecm} 
\begin{minipage}[t]{7truecm}
\centering
\includegraphics[width=7cm]{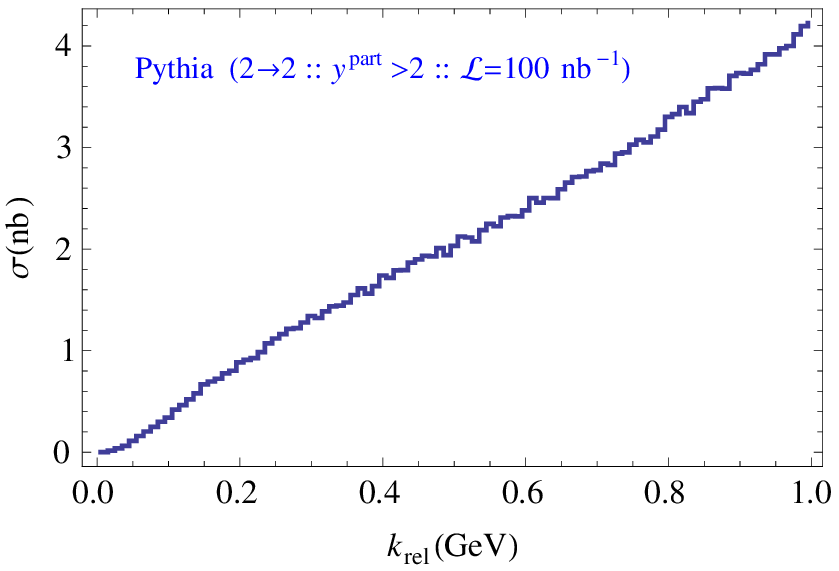}
\caption{ Same plot as in Fig.~\ref{herfullc} but using Pythia. We show these curves in a wide range of $k_{\rm rel}$ to give an idea of the remarkable Monte Carlo stability against fluctuations  achieved on account of  the very high statistics used.  
}
\label{pythfull}
\end{minipage}
\end{figure}

We have used Herwig and Pythia to compute hadron final states from  $2\to 2$ QCD parton processes reaching a Monte Carlo luminosity  
${\cal L}\sim 100$~nb$^{-1}$.  In Fig.~\ref{herfullc} we show the integrated cross section as a function of the center of mass relative momentum
in the $\mol$ molecule obtained using Herwig. To get the minimal experimental value of $\sigma\sim 3.1\pm 0.7$~nb we need to include $\mol$ configurations having up to
$k_{\rm rel} = 205\pm 20$~MeV.  Molecule candidates in the ball of relative momenta  ${\cal R}$ can account only for $0.071$~nb.
Repeating the same calculation with Pythia, see  Fig.~\ref{pythfull}, we get 
$k_{\rm rel} = 130\pm15$~MeV whereas in ${\cal R}$ we integrate $0.11$~nb.

Simulating the real experimental situation of prompt production of $X(3872)$ at CDF would require a further increase of just a factor of $10^4$ in the Monte Carlo luminosity which is extremely CPU demanding. Yet, in consideration of the stability of our results, 
we do not expect significant variations from what here observed.

In conclusion we study $g c\bar c$ events with one gluon at $p_\perp > 5$~GeV  recoiling from the $c\bar c$ pair which in turn can hadronize into open charm mesons very close in phase space. We perform this computation at the parton level using ALPGEN~\cite{alpgen} and assuming that the fragmentation functions into open charm mesons to be set to one. This corresponds to  an upper bound estimation. The results obtained point at a definitely negligible contribution from these configurations, being in the range of few picobarns.

{\bf \emph{Conclusions}}. 
We have simulated the production of open charm mesons in high energy hadronic collisions at the Tevatron. The generated samples have been examined searching for $D$ and $D^*$ mesons being in the conditions to form, through resonant scattering, bound states with  binding energy
as small as $\sim 0.25$~MeV. These $X(3872)$ candidates have been required to pass the same 
kinematical selection cuts used in the CDF data analysis. This allows to estimate an upper bound for the theoretical prompt production cross section of $X(3872)$ at CDF. Averaging the results obtained with Pythia and Herwig we find this to be approximately $0.085$~nb in the most reasonable region of center of mass relative momenta $[0,35]$~MeV of the open charm meson pair constituting the  molecule.  This value has to be compared with the lower bound on the experimental cross section, namely $3.1\pm 0.7$~nb, extracted from CDF data
The intuitive expectation that $S-$wave resonant scattering is unlikely to allow the formation of a loosely bound $\mol$ molecule in high energy hadron collision is confirmed by this analysis.

\begin{acknowledgments}
We wish to thank Marco Rescigno for his indispensable hints on CDF data. We acknowledge many useful discussions with  E. Braaten, R. Escribano, F. Maltoni, R.L. Jaffe,  and thank Gino Isidori and Alessandro Strumia for their comments on the manuscript.
The work of one of us (B.G.) is supported in part by the US Department of Energy under contract DE-FG03-97ER40546.

\end{acknowledgments}

\bigskip 

\end{document}